\documentclass[12pt]{article}
\usepackage{amssymb}
\usepackage{epsfig}
\usepackage{graphicx}
\usepackage{amsmath}
\usepackage{verbatim}

\csname @addtoreset\endcsname{equation}{section}

\def \be  {\begin{equation}}
\def \ee  {\end{equation}}
\def \ba  {\begin{eqnarray}}
\def \ea  {\end{eqnarray}}

\begin{document}
\begin{titlepage}
\title{\bf\Large The Goldstino Field in Linear and Nonlinear Realizations of Supersymmetry  \vspace{18pt}}

\author{\normalsize Hui~Luo,~Mingxing~Luo and Liucheng~Wang  \vspace{12pt}\\
{\it\small Zhejiang Institute of Modern Physics, Department of Physics,}\\
{\it\small Zhejiang University, Hangzhou 310027, P.R. China}\\
}

\date{}
\maketitle \voffset -.3in \vskip 1.cm \centerline{\bf Abstract}
\vskip .3cm
A Goldstino field in the nonlinear realization of supersymmetry is constructed from an appropriate chiral super-multiplet of the linear theory,
in general O'Raifeataigh-like models.
The linear theories can thus be reformulated into their nonlinear versions, via the standard procedure.
The Goldstino field disappears totally from the original Lagrangian in the process, but reemerges in the Jacobian of the transformation and covariant derivatives.
Vertices with Goldstino fields carry at least one space-time derivative, as one would have expected.

\vskip 5.cm \noindent November  2009
 \thispagestyle{empty}

\end{titlepage}
\newpage

In general O'Raifeataigh-like (OR) models, it is convenient to start with the linear realization of supersymmetry (SUSY) \cite{OR}.
For the discussion of spontaneously breaking of SUSY in these models, only chiral super-multiplets need to be concerned, be they elementary or composite.
Every chiral super-multiplet $\Phi$ has three component fields $(\phi,\psi,F)$, which transform into each other via
(see, for example, \cite{WB})
\begin{eqnarray}
 \delta_{\xi} \phi & = & \sqrt 2 \xi \psi,  \nonumber \\
 \delta_{\xi} \psi_\alpha &=& \sqrt{2} F  \xi_\alpha+ i \sqrt 2 (\sigma^\mu \bar \xi)_\alpha \partial_\mu \phi~, \label{lsusy} \\
 \delta_{\xi} F &=& i \sqrt 2 \bar \xi \bar \sigma^\mu \partial_\mu \psi . \nonumber
\end{eqnarray}
When the $F$-term of one chiral super-multiplet $\Phi_0$ develops a non-zero vacuum expectation value\footnote
{We will always assume that only the $F$-term of $\Phi_0$ has a non-zero VEV,
to simplify presentation.
In case that several super-multiplets have non-zero VEVs,
the super-multiplets can always be realigned, such that only $\Phi_0$ will have a non-zero VEV.} (VEV) $\langle F_0 \rangle$,
SUSY is spontaneous broken.
The corresponding $\psi$ field will transform inhomogeneously under SUSY transformations via
\be \delta_\xi \psi_{0\alpha} =\sqrt2 \langle F_0 \rangle \xi_\alpha + \cdots
\label{chiral0}
\ee
According to the general theory of spontaneously symmetry breaking, $\psi_0$ will have a zero mass and is
usually referred to as the Goldstino field in the linear theory.

To deal with the low energy physics of theories with spontaneously symmetry breaking,
it is usually advantageous to work with their nonlinearly realized versions.
It can be particularly useful if the system is strongly coupled, as exemplified by the low energy effective theory of hadronic physics.
In the standard realization of nonlinear SUSY,
a Goldstino field $\tilde\lambda$ is presumed to exist and to transform as \cite{VA},
\begin{eqnarray}{\label{standard1}}
\delta_\xi \tilde{\lambda}_{\alpha}
&=&\frac{\xi_{\alpha}}{\kappa}- i \kappa (\tilde{\lambda}\sigma^\mu \bar{\xi} -\xi\sigma^\mu\bar{\tilde{\lambda}})
\partial_{\mu}\tilde{\lambda}_{\alpha}
\ea
under SUSY transformations.
All other fields will be referred to as matter fields and they are assumed to transform as,
\ba
\delta_\xi \varphi&=&-i\kappa(\tilde{\lambda}\sigma^\mu \bar{\xi} -\xi\sigma^\mu\bar{\tilde{\lambda}})
\partial_{\mu} \varphi
{\label{standard2}}
\end{eqnarray}
In certain circumstances, it could be expedient to define an equivalent $\lambda$, via \cite{Wess83}
\begin{eqnarray}{\label{T}}
\lambda_{\alpha}(x)=\tilde{\lambda}_{\alpha}(z),\ \ \ \
z =x-i\kappa^{2}\tilde{\lambda}(z)\sigma \bar{\tilde{\lambda}}(z)
\end{eqnarray}
such that the Goldstino and matter fields transform as,
\begin{equation}{\label{chiral1}}
\delta_{\xi}\lambda_{\alpha}
=\frac{\xi_{\alpha}}{\kappa}-2i\kappa\lambda\sigma^\mu \bar{\xi}\partial_\mu \lambda_{\alpha}
\end{equation}
\begin{equation}{\label{chiral2}}
\delta_{\xi}\varphi=-2i\kappa\lambda\sigma^\mu \bar{\xi}\partial_\mu \varphi
\end{equation}
The latter version is particularly convenient when dealing with chiral super-multiplets.
It will thus be referred to as the chiral version of nonlinear realization
and adapted in discussions below when the situation warrants.

Obviously, $\lambda$ is closely related, but not exactly identical, to $\psi_0$.
In particular, the transformation rules in (\ref{chiral0}) and (\ref{chiral1}) are not the same, though quite close.
Given the central role played by the $\lambda$-field in the formulation of standard realization of nonlinear SUSY,
an explicit construction of $\lambda$ out of $\Phi_0$ is important by its own right.
In addition, such a construction will be a necessity
when one tries to connect low energy nonlinear supersymmetric effective theories with their ultraviolet completions.

The situation is in parallel with the Goldstone fields in low energy hadronic physics \cite{Weinberg}.
The Lagrangian in the linear $\sigma$-model
$$
\mathcal{L}=-\frac{1}{2}\sum_{n=1}^4\partial_{\mu}\phi_n\partial^{\mu}\phi_n-\frac{\mathcal{M}^2}{2}\sum_{n=1}^4\phi_n\phi_n
-{g\over4}\sum_{n=1}^4(\phi_n\phi_n)^2 \label{lsigma}
$$
is invariant under $SO(4)$-rotations among $\phi_n = (\vec\phi, \phi_4)$.
In particular, $\mathcal{L}$ is invariant under the axial-isospin subgroup:
\be
\delta\vec{\phi}=2\vec{\epsilon}\phi_4,~~\delta\phi_4=-2\vec{\epsilon}\cdot\vec{\phi}
\ee
When $\mathcal{M}^2<0$, $\phi_4$ develops a non-zero VEV $\langle \phi_4 \rangle$ and the axial-isospin symmetry is broken.
$\vec\phi$ transform inhomogeneously under axial-isospin rotations via
\be
\delta\vec{\phi}=2\vec{\epsilon} \langle \phi_4 \rangle + \cdots
\ee
They have vanishing masses and are the Goldstone fields in the linear version.
In order to calculate scattering amplitudes at low energies,
it is advantageous to convert $\mathcal{L}$ into its nonlinear version
$$
\mathcal{L}^{\rm NL}=-\frac{1}{2}\partial_{\mu}\sigma\partial^{\mu}\sigma - {\sigma^2 \over 2 |\langle \phi_4 \rangle|^2 } \vec{D}_{\mu}\vec{D}^{\mu}
-\frac{1}{2}\mathcal{M}^2\sigma^2-\frac{\lambda}{4}\sigma^4 \label{nlsigma}
$$
by introducing the nonlinear Goldstone fields $\vec\pi = 2 \langle \phi_4 \rangle \vec\zeta$, where
\be
\sigma=\sqrt{\sum_n \phi_n^2}, \ \ \zeta_a = {\phi_a \over \phi_4+\sigma},\ \
\vec{D}_{\mu} = \frac{\partial_{\mu}\vec{\zeta}}{1+\vec{\zeta}^2}
\ee
$\vec \phi$ and $\vec \pi$ are closely related to each other and $\vec \pi  \simeq \vec \phi $ to the leading order.
The virtue of $\mathcal{L}^{\rm NL}$ is that for every term that involves the $\vec\pi$ must contain at least one space-time derivative of the field.
These derivatives introduce factors of the Goldstone boson energy when calculating scattering amplitudes,
which can be used as expansion parameters in low energy processes.
Different from $\vec \phi$, $\vec{\zeta}$ now transform nonlinearly into themselves under axial-isospin rotations
\be
\delta \vec{\zeta}=\vec\epsilon(1-\vec{\zeta}^2)+2\vec{\zeta}(\vec{\epsilon}\cdot\vec{\zeta})
\ee

In a sense, the chiral super-multiplet $\Phi_0$ is the analog of the $\phi_a$'s in the linear $\sigma$ model,
while $\lambda$ is the analog of $\vec \pi$ in the nonlinear version.
In this letter, we shall show how the $\lambda$ field is related to the component fields of $\Phi_0$,
in parallel as how the $\vec \pi$'s are related to the $\phi_a$'s.
It will be evident that $\lambda \simeq \psi_0$ to the first order of approximation, if $\kappa$ is chosen properly.
Guided by this newly found relation, we may reformulate any spontaneously broken linear theories into nonlinear ones,
via the standard procedure developed in \cite{IK78,IK77,IK82}.
In the process, the Goldstino field disappears totally from the original Lagrangian,
but reemerges in the Jacobian of the transformation and covariant derivatives.
Vertices with Goldstino fields carry at least one space-time derivative, in parallel with the nonlinear $\sigma$-model.
Accordingly, the energies of the Goldstino fields can again be used as expansion parameters in low energy processes.
This can be particularly useful when the underlying physics is strongly coupled and the usual perturbation theory fails.
Even if the underlying physics is weakly coupled,
the nonlinear version thus formulated still serves as a good organization device in calculations of scattering amplitudes at low energies.

As illustrated in \cite{IK78,IK77,IK82},
any linear super-multiplet can be prompted to a nonlinear one with the help of $\tilde\lambda$,
\begin{equation}{\label{l-to-nl}}
\Omega^\sigma =\exp\left[-\kappa \left(\tilde\lambda Q+\Bar{\tilde\lambda} \Bar{Q}\right)\right]\times \Omega
\end{equation}
Under SUSY transformations,
\begin{eqnarray}
\delta_{\xi} \Omega^\sigma=-i \kappa(\tilde{\lambda}\sigma^\mu \bar{\xi} -\xi\sigma^\mu\bar{\tilde{\lambda}})\partial_\mu \Omega^\sigma
\end{eqnarray}
That is, all component fields in $\Omega^\sigma$ transform into themselves
and independent of one another.
Any of them can be integrated out without breaking SUSY.
Of course, whether and how to integrate out a component field are dynamical questions.
When one or more component fields have masses much higher than the typical energy scale in the concerned physical process,
it is usually convenient to integrate them out.
Ignoring quantum fluctuations,
one may express these heavy fields in terms of the light ones via the relevant equations of motion.
Effectively, the heavy fields are substituted by a set of high order operators constructed out of the remaining light fields.
When the heavy fields are extremely heavy, these newly introduced high order operators can be ignored.
In that case, the end result can be obtained by simply setting the heavy field to zero.

Now the issue is how to identify the $\tilde\lambda$ in a given model.
In a generic OR model, the Goldstino field resides in a chiral super-multiplet $\Phi_0$,
whose $F$ term has a non-zero VEV, as mentioned above.
Taylor expanding $\Phi_0(y,\theta)=\phi_0(y)+\sqrt 2\theta \psi_0(y)+\theta^2F_0(y)$ ($y = x + i \theta \sigma \bar\theta$).
The corresponding nonlinearly realized super-multiplet can be obtained via (\ref{l-to-nl})
\be
\Phi^\sigma_0 = \Phi_0(y - 2 i \kappa \theta \sigma \bar{\tilde\lambda}(x) + i \kappa^2 \tilde\lambda(x) \sigma \bar{\tilde\lambda}(x), \theta - \kappa \tilde\lambda(x))
\label{phisig}
\ee

We now propose to construct $\tilde\lambda$ out of the components of $\Phi_0$,
by demanding $\psi^\sigma_0$ to vanish.
The algebra simplifies significantly, if one re-expresses $\tilde\lambda$ in terms of $\lambda$ via (\ref{T}).
Explicitly, one gets\footnote{
For any chiral super-multiplet without a VEV, one can also construct a $\lambda$ with the transformation property (\ref{chiral1}) via (\ref{keyrel}).
But such a construction is of no practical use,
as it cannot be used to separate the Goldstino field from the others.} by setting $\psi^\sigma_0=0$,
\be
\lambda = {\psi_0 \over \sqrt 2\kappa F_0} - i {\sigma^\mu\Bar \lambda^{'} \over F_0 }
\left( \partial_\mu  \phi_0 - \sqrt 2 \kappa \lambda \partial_\mu\psi_0 +  \kappa^2 \lambda^2 \partial_\mu F_0 \right)
\label{keyrel}
\ee
where
\be
\bar\lambda^{'}= \bar\lambda - 2 i\kappa^2\lambda \sigma^\mu \bar{\lambda} \partial_\mu \bar\lambda
- 2 \kappa^4 \lambda^2  \bar{\lambda} \bar \sigma^\nu\sigma^\mu \partial_\nu \bar{\lambda}\partial_\mu \bar\lambda
+ \kappa^4 \lambda^2 \bar\lambda^2  \partial^2 \bar\lambda
\ee
This relation can serve as a definition of $\lambda$, from which the nonlinear Goldstino field is constructed out of $\phi_0$, $\psi_0$, and $F_0$.
Taking $\kappa^{-1} =\sqrt2 \langle F_0 \rangle$, one has $\lambda\simeq\psi_0$ to the leading order of approximation.
It is straightforward to check that the $\lambda$ defined as such does indeed transform according to (\ref{chiral1}) as the linear fields transform according to (\ref{lsusy}).
Since both $\lambda$ and $\psi_0$ are Grassmann variables, a closed (though complicated) form of $\lambda$ in terms of the linear fields can be obtained from (\ref{keyrel}) by iteration.
But such a closed form of $\lambda$ is not needed in practice.
We need (\ref{keyrel}) only to show that a nonlinear $\lambda$ with the desired transformation property (\ref{chiral1})
can be constructed out of the linear ones reside in $\Phi_0$.
More importantly, $\psi^\sigma_0$  vanishes by definition when such a $\lambda$ is used in the conversion (\ref{l-to-nl}).

We emphasize again that the rationale for setting any heavy field to zero is due to dynamics.
It happens when the field is extremely heavy compared with the typical energy scale of the concerned physical process.
It decouples from all of the low energy physics.
On the other hand, $\psi_0$ is not heavy and is actually massless.
It cannot be dropped by the reasoning of decoupling.
$\psi^\sigma_0=0$ is realized by an appropriate definition of $\lambda$.
Its feasibility is due to the SUSY algebras in (\ref{lsusy}) and (\ref{chiral1}).
As advertised above, (\ref{keyrel}) is the analog of representing the $\vec \pi$ fields in the nonlinear $\sigma$-model in terms of the $\phi_a$ fields in the linear $\sigma$-model \cite{Weinberg}.

In \cite{Seiberg}, a chiral super-multiplet with the constraint $X_{NL}^2=0$ is proposed to describe the Goldstino.
In that case, (\ref{keyrel}) simplifies into $\lambda =\psi_{NL} / \sqrt 2\kappa F_{NL}$ \cite{LLZ}.
More so, one has now $X_{NL}^\sigma=\theta^2 F^\sigma$.
That is, the Goldstino component disappears in the corresponding nonlinearly realized super-multiplet, as it should be.

We are now in the position to reformulate any linear theory into the corresponding nonlinear one, via the standard procedure presented in \cite{IK78,IK77,IK82}.
To work out the Lagrangian, it is actually convenient to use the non-chiral version of nonlinear realization.
We need to define the $\lambda$ via (\ref{keyrel}) first and obtain $\tilde \lambda$ afterward via (\ref{T}).
With this $\tilde \lambda$, we change all linear super-multiplets into their nonlinear version via (\ref{l-to-nl}).
This process eliminates $\psi^\sigma_0$ automatically in $\Phi^\sigma_0$.
So the Goldstino field disappears in the original Lagrangian, but reemerges in the Jacobian of the transformation and covariant derivatives.
Vertices with Goldstino fields carry at least one space-time derivative.
It is reassuring that the salient features of nonlinear theories are all retained.

In particular, the Goldstino field is totally absent in all potential terms,
where space-time derivatives are not allowed.
Actually, all potential terms in the nonlinear version are included in
\be
\int d^4x (d^2 \theta W(\hat\Phi^\sigma) + h.c.) \label{wn}
\ee
where all $\hat\Phi^\sigma$ are obtained from $\Phi^\sigma$ by setting $y$ to $x$ in (\ref{phisig}).
It has the same structure with the linear version
\be
\int d^4x (d^2 \theta W(\Phi) + h.c.) \label{wl}
\ee
Since $\psi_0$ is massless, there cannot be bilinear terms $\psi_0\psi_0$ or $\psi_0 \psi_i$ in (\ref{wl}) to start with.
Thus, the mass spectrum of the model is not changed by going from the linear version to nonlinear one by setting $\psi_0^\sigma=0$.

In this nonlinearly realized version, all heavy fields are kept.
Depending on the specific physical processes concerned, one may integrate out part or all of them according to the mass spectra,
with the help of relevant equations of motion.

In principle, the above discussion can be extended to Fayet-Iliopoulos models.
In those cases, the Abelian gauge field is described by a real super-multiplet $V= D \theta^2 \bar\theta^2 + \chi \theta \bar\theta^2 + \bar\chi \bar\theta \theta^2 + \cdots$.
When the so-called Fayet-Iliopoulos term is present, the $D$-field in $V$ acquires a non-zero VEV and SUSY is spontaneously broken.
Correspondingly, the $\chi$ field is massless and plays the role of the Goldstino field.
A nonlinearly realized super-multiplet $V^\sigma$ can again be obtained via (\ref{l-to-nl}).
The nonlinear Goldstino field $\tilde\lambda$ can be defined in terms of the component fields in $V$ by demanding $\chi^\sigma=0$.
However, there could be potentially subtle issues related to gauge invariance.
On the other hand, there are strong arguments against the presence of Fayet-Iliopoulos terms,
in the context of supergravity \cite{Seiberg1,Dienes}.
These issues will not be addressed in this letter and should be dealt with in future works.

The Goldstino field $\tilde\lambda$ itself can also be
prompted to a linear super-multiplet \cite{WB} by using the SUSY transformation in (\ref{standard1}),
\[ 
\Lambda(\tilde\lambda)=\exp(\theta Q+\bar{\theta} \bar{Q})\times\tilde\lambda
\] 
so are all matter fields by using the SUSY transformation in (\ref{standard2}),
\[ 
\Xi(\tilde\lambda)=\exp(\theta Q+\bar{\theta} \bar{Q})\times\varphi
\] 

From an arbitrary SUSY non-invariant action,
\[
S_0 = \int d^4 x L(\partial_\mu \varphi, \varphi)
\]
one can easily construct a SUSY invariant one, with the help of $\Lambda$ and $\Xi$,
\be
S = \kappa^4 \int d^4x d^4\theta \Lambda \Lambda \bar\Lambda \bar\Lambda L(\partial_\mu \Xi, \Xi).
\ee
Notice that $\Lambda(x) =\kappa^{-1}\theta^{'} = \kappa^{-1}\theta +\tilde \lambda(z)$, and $\Xi(x) = \varphi(z)$,
where $z=x - i \kappa \tilde \lambda(z) \sigma \bar \theta +i \kappa \theta \sigma \bar{\tilde \lambda}(z)$.
The integration over the Grassmann variables can be carried out by changing integration variables from $(x,\theta)$ to $(z, \theta^{'})$.
A straightforward calculation yields,
\be
S = \int d^4 x \det (T) L(\nabla_\mu \varphi, \varphi)
\ee
where $T_\mu^\nu = \delta_\mu^\nu - i\kappa^2 \partial_\mu \tilde \lambda\sigma^\nu \bar{\tilde \lambda}
+ i \kappa^2\tilde\lambda \sigma^\nu \partial_\mu \bar {\tilde \lambda}$.
and $\nabla_\mu = (T^{-1})_\mu^\nu \partial_\nu$.
This action can also be obtained from $S_0$ by simply prompting $\partial_\mu$ to $\nabla_\mu$ in $L$,
and inserting the determinate of matrix $T_\mu^\nu$ \cite{CL04,CL96}.
One may check that $S$ is invariant under the combined transformations of (\ref{standard1}) and (\ref{standard2}).
Obviously, integrations over the Grassmann variables can always be carried out in a similar manner for arbitrary functionals of $\Lambda$ and $\Xi$.

\section*{Acknowledgement}
We would like to thank Yihong Gao and Guohuai Zhu for valuable discussions.
This work is supported in part by the National Science Foundation
of China (10425525) and (10875103).

\end{document}